# Seamless Data Migration between Database Schemas with DAMI-Framework: An Empirical Study on Developer Experience

Delfina Ramos-Vidal
Alejandro Cortiñas
delfina.ramos@udc.es
Universidade da Coruña
Centro de Investigación CITIC,
Database Lab.
A Coruña, Spain

Miguel R. Luaces
Oscar Pedreira
Ángeles Saavedra Places
Universidade da Coruña
Centro de Investigación CITIC,
Database Lab.
A Coruña, Spain

Wesley K. G. Assunção
North Carolina State University,
Department of Computer Science
Raleigh, United States

## Abstract

Many businesses depend on legacy systems, which often use outdated technology that complicates maintenance and updates. Therefore, software modernization is essential, particularly data migration between different database schemas. Established methodologies, like model transformation and ETL tools, facilitate this migration. However, these methodologies require deep knowledge of database languages and both the source and target schemas. This necessity renders data migration an error-prone and cognitively demanding task. Our objective is to alleviate developers' workloads during schema evolution by proposing DAMI-Framework. This framework incorporates a domain-specific language (DSL) and a parser to facilitate data migration between database schemas. DAMI-DSL simplifies schema mapping, while the parser automates SQL script generation. To evaluate DAMI-Framework, we conducted an empirical evaluation with 21 developers to assess their experiences using our DSL versus traditional SQL. The study allows us to measure their perceptions of the DSL properties and user experience. The participants praised DAMI-DSL for its readability and ease of use. The findings indicate that our framework has positive impact in data migration efforts compared to SQL scripts, by reducing the lines of code and characters required for the migration by 18.3% and 35.6%, respectively.

## 1 Introduction

Most existing industrial software is long-lived and represents several years of business value [19]. Throughout the software life cycle, the architecture decays, user requirements change, and technologies evolve, leading to legacy systems [8]. Legacy systems are costly to maintain, more exposed to cybersecurity risks, less effective in meeting their intended purpose, and increase the costs of digital transformation [7, 14, 25]. Thus, companies must modernize their legacy systems to remain competitive, preserving the valuable knowledge gained through years of system development [3, 19, 41]. As part of modernization, a crucial step is data migration, which aims to move the data from the legacy database into the modern system. Data migration becomes complex because the source and destination databases are commonly structurally different [24]. Furthermore, during data migration, the data must be consistent across both data sources, which is also a challenging task [11].

A standard solution to data migration is to use ETL (Extract, Transform, and Load) techniques. Several ETL tools have emerged in recent years [36]. However, while these tools are practical, they typically fail to offer a comprehensive and intuitive answer [9, 31]. Some are too complex, requiring substantial training and experience, while others are proprietary, burdening organizations with licensing restrictions and dependencies and even hiring and retaining specialized developers [20]. Commercial ETL solutions capable of performing intricate mapping and transformation operations are too complicated, technically demanding, and require a significant upfront investment, even when that level of sophistication is not always necessary [9]. In addition to the limitations identified in the literature, our extensive practical experience—spanning over two decades of software development and legacy system maintenance—has exposed us to two common challenges in data migration. These challenges are determining *what to migrate* (i.e., mappings between the source and target schemas) and *how to migrate* (i.e., consistently retrieving and transforming data conforming with constraints and structure of the target schema).

In response to these challenges, we present a solution specifically designed for **Da**ta **Mi**gration, called DAMI-Framework. Our framework consists of two key components: (i) a Domain-Specific language (DSL) designed to handle the complexities of data migration during the modernization of legacy software, defining the mapping and transformations required to migrate the data from a legacy system to a new system, which commonly has different database schemas; and (ii) a DSL Parser that validates migration scripts written with our DSL and auto-generates the SQL scripts required to complete the migration operations. We designed DAMI-Framework to be flexible and customized precisely for the task, focusing on seamlessly moving data from the prior legacy database schema to the new one.

To evaluate DAMI-Framework, we present an empirical study based on a real scenario faced by engineers to migrate the webpage of a scientific repository. In this study, we detail both the source and target database schemas, as well as the migration challenges addressed using our DSL and parser. By applying DAMI-Framework, we reduced the lines of code and characters required for the migration by 18.3% and 35.6%, respectively. The evaluation included an experiment with 21 software developers who addressed a data migration challenge. This segment of the evaluation focuses on user experiences, particularly assessing the ease of learning and using our solution, the efficiency of the migration process, and overall user satisfaction. Based on our real-life scenario, we designed a task for the study participants to perform. From a questionnaire survey composed of 12 questions administered to participants after they completed the task, we found that software developers generally agree or strongly agree with the benefits provided by the properties of DAMI-DSL, such as efficiency, conciseness, expressiveness, and flexibility. Additionally, they reported a positive user experience, reflected in their feedback regarding confidence, ease of use, readability, and satisfaction. These user insights are valuable for enhancing DAMI-framework, ensuring it addresses the diverse needs of individuals involved in legacy software modernization.

## 2  Motivation and Problem Statement

Data migration is a recurring problem in software development [17, 45, 49]. This problem appears when an information system evolves or when a new information system replaces an existing one with persistent data that must be preserved [5, 35]. Data migration is frequent nowadays because of the need to modernize legacy systems [3]. Modernizing legacy systems involves adapting or re-engineering the source code using new technology and redesigning the database schema (i.e., data model) to ensure that it still meets the domain's demands. Then, strategic data available in the legacy systems must also be migrated to the new database, matching the new schema, leading to two main challenges, as follows.

Defining *what needs to be migrated* is the first challenge when upgrading a database. This initial step involves understanding the structure (i.e., schema or model) and content (i.e., data) of the legacy database (*source*) and determining how it maps to the new schema (*target*) of the modern systems. Engineers responsible for data migration must examine the legacy schema to determine which tables, columns, constraints, and relationships must be mapped. Analyzing the legacy database is not only a laborious task, as even simple systems have dozens of tables, columns, constraints, and relationships, but it is also a cumbersome activity since legacy systems are usually poorly documented, with degraded database schemas due to several years of unplanned maintenance. In summary, the first challenge on data migration is: *Challenge 1 - identifying the mapping between the source schema and the target schema*.

Let us assume that engineers could deal with the complexity of identifying the mappings. However, data migration is still needed. Then, the second challenge is *how to transform the data*, which requires careful consideration. Transforming data from a legacy database schema to a new one usually requires writing complex SQL queries (e.g., "insert" statements depending on "select" statements with multiple tables connected by "join" clauses) to merge or split data instances, including mapping source tables to their corresponding target tables and deciding how to transform the data for compatibility. Legacy databases often include duplicate, inconsistent, or poorly organized data due to years of use. Addressing these issues ensures the new schema maintains high data quality standards. During data migration, we must ensure that the data types are compatible and preserve necessary constraints (such as primary keys, foreign keys, and unique constraints). Preserving the dependencies and relationships between tables is crucial, as adequately managing them is essential for maintaining data integrity and consistency. In summary, the second challenge on data migration is: *Challenge 2 - retrieving and transforming consistent data from the legacy database while conforming to the constraints and structure of the modern database schema*.

To illustrate the two challenges described above, we use a real data migration case for the website of a scientific repository. This website serves as a central repository for scientific publications and details of affiliated researchers. However, the technology stack has become outdated since its launch in the early 2000s. This required us to modernize the website to enhance its efficiency and maintainability, especially as new developers joining the team are not familiar with the legacy technologies, and the legacy website has an over-engineered structure.

Figure 1 illustrates the legacy and modern websites' database schema. Note that although we describe a real-world scenario, we restricted the use case to publication management due to space constraints. Both schemas were designed to store information about various types of scientific publications (journals, conferences, books, and book chapters) and their country of publication. Additionally, they maintain a list of researchers and track authorship details of each published paper. The legacy database schema (left side of the figure) suffers from inefficiencies, particularly data duplication. The legacy database models scientific publications using an inheritance strategy where PUBLICATION is the supertype, and JOURNAL, CONFERENCE, BOOK, and BOOK_CHAPTER are the subtypes. For instance, if researchers publish two papers at the same conference, there would be duplicate entries—one for each paper in PUBLICATION and two for the same conference in CONFERENCE, leading to redundancy.

Engineers restructured the database schema to eliminate data duplication as part of the modernization. To do so, they dissolved the inheritance structure of the database and separated publication details from paper information. The new schema includes two main tables: PAPER, which stores details about each paper, and PUBLICATION, which holds information about the journal, conference, book, or book chapter where the paper was published. A new table, PUBLICATIONTYPE, categorizes the different publication types and relates to PUBLICATION. The country of publication, previously in a separate table, is now an attribute in PUBLICATION. We also created a new PUBLISHER table to collect details about publishing houses.

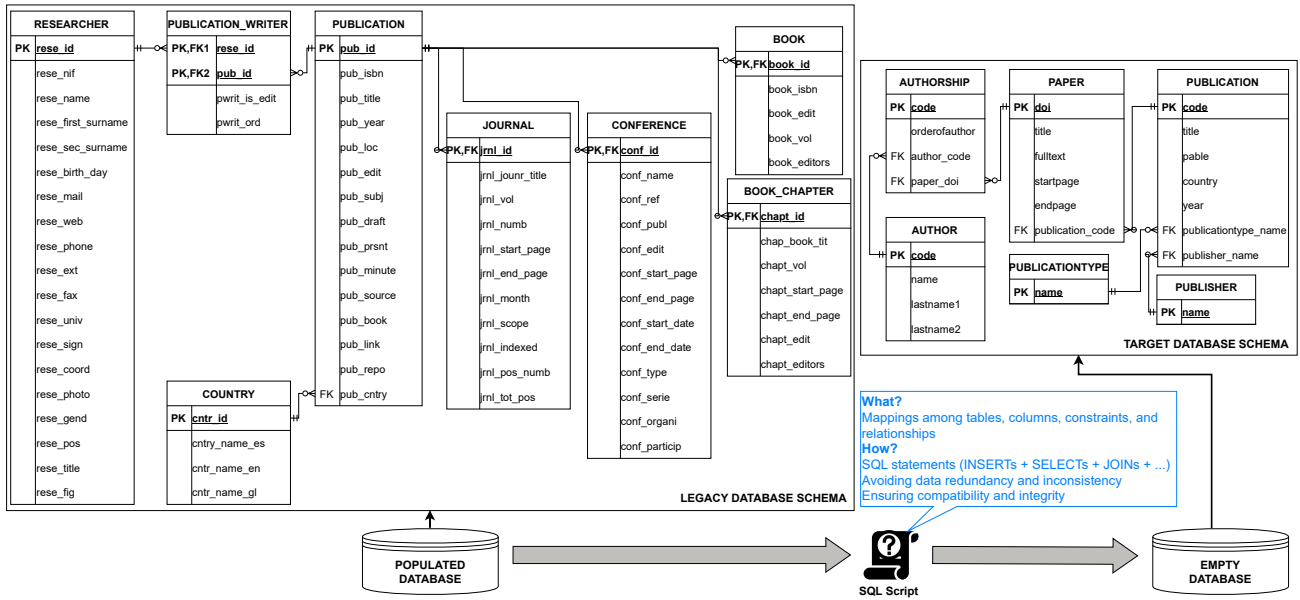

Figure 1: Database schemas (populated legacy database and empty target) for modernizing the scientific repository website and challenges of data migration.

This change helps avoid duplication since the same publisher can manage multiple journals and books.

The migration process needs careful analysis to identify tables, columns, and relationships since no direct mapping exists between the old and new databases. For instance, tables PUBLICATION, JOURNAL, CONFERENCE, BOOK, BOOK_CHAPTER, and COUNTRY must be mapped to PAPER, PUBLICATION, PUBLICATIONTYPE, and PUBLISHER, which demands input from a domain expert to define the mappings. Once they identify the mappings, the engineers define the transformation of legacy data to fit into the target schema while ensuring data integrity. Engineers also determine which primary keys (PK) to retain or regenerate. For instance, while the PK for PUBLICATION will remain unchanged, the PK for PAPER will be auto-generated. Consequently, there will be a requirement to establish equivalences for the corresponding foreign keys (FK).

In this context, the goal is to define the data migration that will take the data from the legacy database and change it to conform to the target database schema, allowing it to populate the target database. This migration is not trivial, since it includes dissolving an inheritance structure and combining the data from five legacy tables into two tables, while ensuring no duplicates exist.

### 2.1 Limitations of Existing Solutions

Traditional ETL tools, commonly used for data migration, have notable limitations [9]. They require extensive configuration and customization, offer limited transformation flexibility, struggle with scalability, and incur high licensing and maintenance costs [31]. Cloud-based migration services offer scalability and ease of use but have limitations. Google Cloud Data Fusion[1] provides code-free data integration but requires technical knowledge. It supports various database formats but stores data only in Google Data Cloud. It cannot perform data transformations natively, needing additional modules like Google Data Wrangler,[2] which also has limitations and requires separate installation. AWS Database Migration Service (AWS DMS)[3] supports fundamental transformations and connects to multiple databases but only allows one transformation rule per object, supports limited transformation actions, and complex schema changes are not supported. Microsoft Azure Data Factory[4] requires data to be within a network, not on local machines. Migrating PostgreSQL tables from Azure involves working with JSON definitions and custom SQL queries. It only supports basic activities like Copy and Lookup. As for academic solutions, Curino et al. [10] introduced a Schema Modification Operators (SMO) language for complex schema changes. Despite supporting various table and column operations, it lacks control over primary and foreign key management, which is essential for many cases in our framework design, as illustrated by our problem statement.

## 3 DAMI-Framework: A Data Migration Solution

To address the limitations of existing tools and address the posed challenges, we propose DAMI-Framework. Figure 2 illustrates the workflow of DAMI-Framework. Engineers input a *Data Migration Script* that adheres to DAMI-DSL syntax, described in detail below. This script defines the data migration tasks based on the source and target database schemas. DAMI-Framework does not alter or transform the database structure; instead, it uses the *Source*

---
[1]https://cloud.google.com/data-fusion
[2]https://cloud.google.com/data-fusion/docs/concepts/wrangler-overview
[3]https://aws.amazon.com/dms/
[4]https://azure.microsoft.com/en-us/products/data-factory

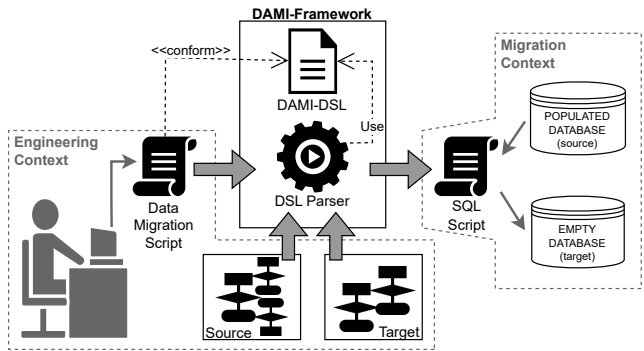

**Figure 2: Overview of the DAMI-Framework**

*Schema*, the *Target Schema*, and the *Data Migration Script* to facilitate the data migration process. In addition to the data migration DSL, DAMI-Framework also includes a *DSL Parser* that processes the *Data Migration Script* and generates an equivalent *SQL script* ready to run against the database.

### 3.1 The DAMI-DSL Metamodel

To use DAMI-Framework, engineers must create a Data Migration Script based on DAMI-DSL, the DSL to define the mappings and transformations needed to migrate the data model from a legacy system to a new target system. DAMI-DSL (see the metamodel in Figure 3) provides a concise and expressive syntax that allows developers to specify data transformations clearly and intuitively. The DSL abstracts the complexities of low-level data manipulation, enabling developers to focus on high-level transformation logic. DAMI-DSL is a declarative language composed of statements that express the mapping from an existing data model to a new target data model without specific implementation details or control flow information. It supports organizing and reusing transformation rules, which improves code maintainability and reduces redundancy in transformation specifications. Additionally, DAMI-DSL includes explicit constructs for common data manipulation operations, such as mapping, empowering developers to express complex transformation requirements in a concise and efficient manner.

Figure 3 depicts the metamodel for DAMI-DSL. This metamodel serves as a formal specification of the syntax, semantics, and structure of DAMI-DSL. It defines the elements and relationships that constitute valid DSL programs, laying the groundwork for tooling support and validation. As previously mentioned, DAMI-DSL supports modularization of the migration rules. The structure of the DSL consists of sentences or statements that define different migration operations, including creation, mapping, insertion, and updating, among others. To better understand the inner workings of the DSL, we provide a brief overview of the operations implemented and currently available in the DAMI-DSL grammar.

- **CREATE**: Used to define the creation of new objects during the migration process. It includes:
  - **CREATE PRODUCT**: Initialize the migration by specifying the product (or project) that embraces the schemas, tables, and attributes.
  - **CREATE CONNECTION**: Establish a connection to source and target databases, specifying database names, hosts, ports, and credentials.
  - **CREATE SCHEMA**: Create new schemas in the target database.
- **MAP**: The most common operation used to map data from legacy tables to target tables. It supports:
  - **MAP original_table TO target_table**: Map individual attributes from legacy to target tables. Multiple mappings can exist, and legacy attributes can be transformed or excluded.
    - **PRIMARY KEY IDENTIFIED WITH**: Map the primary key from the legacy table to the target table, ensuring equivalence between legacy and new identifiers. An auxiliary table stores these mappings.
    - **FOREIGN KEY TO target_table IDENTIFIED WITH original_table.foreign_key**: Associate rows in the legacy table with rows in target table(s), ensuring the mapping of foreign keys to maintain relationships between tables.
    - **UPDATE FOREIGN KEY target_table.foreign_key**: Update foreign key columns in the target table when the required data has already been mapped. If the update fails, the DSL generates an error.
    - **UPDATE FOREIGN TABLE target_table**: Similar to the foreign key update, but used for tables that represent *many-to-many* relationships.
    - **SQL**: Allows the use of SQL native functions for transformations. For example, the statement `SQL: split_part(original_attribute,'-',1) TO target_attribute` splits a legacy attribute and maps the result.
  - **MAP ALL PROPERTIES original_table TO target_table**: Automatically map all attributes with matching names between the legacy and target tables unless otherwise specified. Useful for simple migrations with no transformations.
- **ATTRIBUTE original_table(attribute) TO target_table**: Similar to MAP, but focuses on mapping a single attribute from a legacy table to populate a new target table. It also supports transformation operations.
- **INSERT INTO target_table**: Allows inserting literal values or retrieving data from a legacy table to populate a target table.
- **UPDATE**: Modify existing records in target tables based on certain conditions. Used within a MAP operation.
- **IDENTIFY original_table**: Used to break down join tables, it handles *many-to-many* relationships in legacy systems by transforming them into *one-to-many* relationships in the target system. Combine it with other operations like PRIMARY KEY and FOREIGN KEY to ensure proper data mapping.
- **DROP**: Used to clean up after migration by removing temporary structures and closing connections.
  - **DROP CONNECTION**: Closes database connections.
  - **DROP SCHEMA**: Remove temporary data structures created during the migration.
- **GENERATE SCRIPT**: The final step of the migration, generate the SQL script based on all the specified mappings and transformations and ensure the removal of temporary structures and the termination of database connections.

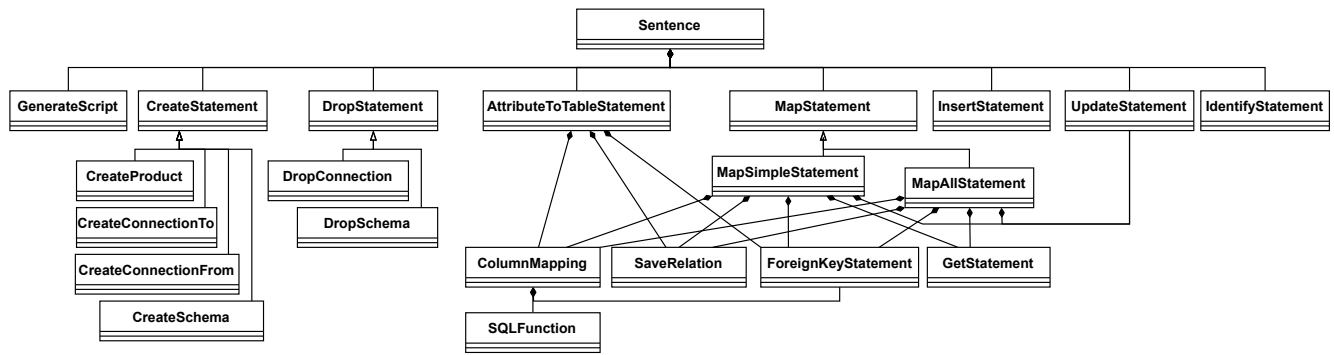

Figure 3: The Metamodel for DAMI-DSL

> **Summary of Key Concepts**
> - DAMI-Framework simplifies data migration with high-level constructs closer to natural language.
> - Modular rules allow reuse and better organization of transformations.
> - Main operations include creating database objects, mapping data, updating records, handling relationships, and cleaning up after the process.
> - The final script generation step outputs the SQL migration script and handles cleanup.

## 3.2 Implementation Aspects

Several tools are available for implementing DSLs [26], such as Lex/Yacc and ANTLR. These tools give developers complete control over the DSL, from syntax to execution style. Our implementation is based on ANTLR [29, 30] to generate the DSL for the DAMI-Framework parser. ANTLR is a powerful parser generator that processes structured text and allows the integration of native code while processing grammar rules.

The main goal of DAMI-DSL is to facilitate data migration between relational databases. To make it user-friendly for developers familiar with SQL, we designed the DSL syntax to resemble SQL, minimizing the user learning curve. ANTLR parses the DAMI-DSL script and triggers a visitor for each statement. The visitor manages the tables and attributes defined in the statement, ensuring they comply with the database schema. Once validated, the visitor generates the corresponding SQL code for execution, allowing for a seamless translation from DSL to SQL while maintaining schema compliance. Section 4 provides an example of the generated code.

## 4 Proof-of-Concept Study

We conducted a proof-of-concept demonstration to assess the feasibility of DAMI-Framework in a real-world scenario. This demonstration involved modernizing the website repository of scientific publications introduced in Section 2. Due to space constraints, we provide only code snippets to illustrate the syntax and usage of DAMI-DSL rather than showing the complete Data Migration Script.

The first step in the migration process involved defining connections to both the legacy and target databases using DAMI-DSL, as demonstrated in Listing 1, including the database name, the host location (e.g., localhost), the connection port, the user with the necessary access privileges, and the corresponding password and schema from which the data will be retrieved. To protect privacy, the example uses mock-up names. The DSL parser processes the data migration script (Listing 1), and generates the corresponding SQL code (Listing 2). This SQL script can be executed on PostgreSQL to establish a connection between the databases. It creates a wrapper with a temporary schema named "*legacy*" for data import, which is deleted after migration, and an auxiliary schema "*aux*" to manage temporary data during the process.

```
1 CREATE CONNECTION FROM (dbname dami_db, host chronos, port
      5432, user dami, pwd dami, schema legacy);
2 CREATE CONNECTION TO (dbname dami_db, host chronos, port
      5432, user dami, pwd dami, schema target);
```

**Listing 1: Snippet for database connection using DAMI-DSL**

```
1 CREATE EXTENSION IF NOT EXISTS postgres_fdw;
2 CREATE SERVER dami_db_database_server FOREIGN DATA WRAPPER
      postgres_fdw OPTIONS (host 'chronos', dbname
      'dami_db', port '5432');
3 CREATE USER MAPPING FOR CURRENT_USER SERVER
      dami_db_database_server OPTIONS (user 'dami', password
      'dami');
4 CREATE SCHEMA legacy;
5 IMPORT FOREIGN SCHEMA legacy FROM SERVER
      dami_db_database_server INTO legacy;
6 CREATE SCHEMA IF NOT EXISTS target AUTHORIZATION dami;
7 CREATE SCHEMA IF NOT EXISTS aux AUTHORIZATION dami;
```

**Listing 2: SQL code for database connection generated by DAMI-Framework**

As described in Section 2, the legacy database stored scientific publications in five tables using an inheritance structure. Listing 3 illustrates how to map data from the legacy tables (`Publication`, `Conference`, and `Country`) to the target table `Publication`. The join condition for these three tables is defined on Line 9. Each column mapping, from Line 2 to 7, specifies the equivalence between legacy attributes and their corresponding attributes in the target table. For instance, Line 5 defines that the `publicationtype_name` column should be populated with the literal value 'CONFERENCE'. Lastly, since the legacy PK `pub_id` will not be preserved, the script ensures that the equivalence between `pub_id` and the newly generated PK (`code`) is recorded, thus maintaining data integrity for subsequent FK migrations. The FK for the publisher is represented as a string (the publisher's name in `pub_edit` from the legacy database). To ensure referential integrity, the `publisher` column in the target table must be pre-populated with all distinct `pub_edit` values. This pre-population step is omitted due to space constraints.

The SQL script generated after parsing the data migration script (Listing 3) is provided in Listing 4. From lines 1 to 5, the script creates an auxiliary structure to store the PK equivalence. Lines 7 to 10 are responsible for migrating and transforming data from the legacy database to the target database. The target table, `Publication`, automatically generates a surrogate identifier using sequences, which demands a solution to maintain the equivalence between the legacy PK and the new PK. To address this, lines 12 to 14 record the relationship between the legacy PK and the new PK, while line 15 cleans up the target database of temporary data.

```
1  MAP publication,conference,country TO publication (
2      conf_name TO title,
3      pub_loc TO place,
4      pub_year TO year,
5      'CONFERENCE' TO publicationtype_name,
6      pub_edit TO publisher_name,
7      cntry_name_es TO country,
8      SAVE RELATION publication.pub_id AS id_publication int
           EQUALS publication.code int
9  ) WHERE (pub_id=conf_id AND pub_cntry=cntry_id);
```

**Listing 3: Mapping three tables to one using DAMI-DSL**

```
1  ALTER TABLE target.publication ADD id_publication int;
2  CREATE TABLE IF NOT EXISTS aux.publication(
3      publication_code int,
4      id_publication int);
5  INSERT INTO target.publication(title, place, year,
         publicationtype_name, publisher_name, country,
         id_publication)
6      SELECT conf_name, pub_loc, pub_year, 'CONFERENCE',
         pub_edit, cntry_name_es, pub_id
7      FROM legacy.publication, legacy.conference,
         legacy.country
8      WHERE pub_id=jrnl_id AND pub_cntry=cntry_id;
9  INSERT INTO aux.publication(publication_code,id_publication)
10     SELECT code,id_publication
11     FROM target.publication;
12 ALTER TABLE target.publication DROP id_publication;
```

**Listing 4: SQL code for mapping tables generated by DAMI-Framework**

Table 1 compares the Lines of Code (LOC) of the data migration script against the generated SQL code, which would otherwise have been written by hand. Although LOC is not directly related to usability [32], the literature supports that for DSLs that are translated into other languages, the efficiency of the DSL can be assessed by comparing the LOC between the DSL and the generated codes [16]. While LOC provide some insight into the size of each script, the structure of DAMI-DSL differs from that of a pure SQL script, making this measure less precise. For example, in SQL, all attributes can be defined inline when defining an INSERT clause, allowing a single action to fit on one line of code. In contrast, DAMI-DSL requires a MAP declaration, where each attribute mapping is specified on a separate line for readability, ensuring that the transformations are apparent and the code is understandable. We also include the character count for each script to enhance this comparison.

**Table 1: LOC in DAMI-DSL vs. SQL**

| Lines Of Code (LOC) | | | Character Count | | |
|---|---|---|---|---|---|
| SQL | DSL | Improvement | SQL | DSL | Improvement |
| 104 | 85 | 18.3% | 4 997 | 3 219 | 35.6% |

By comparing the numbers in Table 1, we observe that the data migration script written in DAMI-DSL requires fewer LOC and characters than its SQL counterpart. Specifically, DAMI-DSL achieves an 18.3% reduction in lines of code and a substantial 35.6% reduction in character count. While these results demonstrate the efficiency of our approach, the primary benefit lies in simplifying the data migration process. This impact will be further evaluated through an experiment with developers, as described in the following section.

## 5 Evaluating Developer Experience

Conducting an empirical study is essential to understand and evaluate the usability, user perception, and limitations of DAMI-DSL. Therefore, we carried out an empirical evaluation based on questionnaire surveys to assess the usability of DAMI-DSL and to determine which type of script—DAMI-DSL or SQL—is more appealing to software developers. We used a cross-sectional survey methodology [42] to gather insights into the usability of DAMI-DSL. This survey enabled us to gain a deeper understanding of the participants' experiences over time and provided opportunities for an in-depth exploration of their opinions. For the design of our evaluation, we reference the ACM SIGSOFT guidelines for questionnaire surveys [37].

### 5.1 Study Design

For the evaluation of DAMI-DSL, we created three artifacts: (i) a *guide* that describes the syntax of DAMI-DSL and provides a usage example based on the scenario used for the Proof-of-Concept study; (ii) a *practical evaluation task* in which participants had to perform a migration task using both DAMI-DSL and SQL; and (iii) a *questionnaire survey* that recorded participants' impressions on the DSL itself, and in comparison to a general-purpose language (GPL) such as SQL. All artifacts used for the evaluation are available online [38]. Participants received the DAMI-DSL guide at the beginning of the three-hour survey session. The first 45-60 minutes were dedicated to a presentation of DAMI-DSL, covering its purpose and providing some examples, with an opportunity for participants to ask clarifying questions. Following this, participants engaged in the evaluation task, which involved an excerpt of the data migration process used as a Proof-of-Concept. The preliminary steps of the migration task were provided both in SQL and DAMI-DSL to serve as examples. The final stages of the migration task were left blank, requiring participants to define how data from various sources (BOOK, BOOK_CHAPTER, CONFERENCE, and JOURNAL) should be migrated into a unified table (PAPER) using both SQL and DAMI-DSL.

To minimize bias based on the learning effect [39], participants were divided into two groups: half completed the migration task using DAMI-DSL first and then SQL, while the other half started with SQL and then used DAMI-DSL. Each group was given one hour to complete both tasks, and the time spent on each was recorded individually. After finishing the tasks, participants submitted their solutions, allowing us to identify errors, and shared their impressions of DAMI-DSL through a structured questionnaire survey. Even though the questionnaire was available online, participants were asked to complete all questions before leaving the experiment, ensuring a 100% response rate. The authors monitored all participants throughout the experiment to avoid dropouts or non-responses.

To ensure a thorough and systematic approach to our survey, we implemented the Goal Question Metric (GQM) methodology [6] to design the questionnaire. The GQM methodology has been proven effective in quantifying the usability aspects of DSLs [2, 33]. In line with this methodology, Table 2 outlines the goals and the corresponding questions formulated to address the goals within the usability domain comprehensively. The table also includes the metrics (measurements used to answer the questions) for each question, enabling us to evaluate the defined goals.

The questionnaire (Table 2) incorporates 12 closed-ended quantitative questions. Each question asked engineers to rate a specific aspect within the respective area using a five-point Likert scale [43]. This allowed the participants to express their agreement or disagreement with particular statements regarding the usability, effectiveness, and satisfaction with the DSL.

As per existing literature, DSL evaluation encompasses four key aspects: expressiveness, conciseness, integration, and performance [13]. This paper focuses on the first two aspects, as they hold importance concerning the DSL itself. We aim to assess the specifications necessary for software engineers to comprehend and/or generate the data migration specification, rather than examining the interaction of the DSL with specific tools. We follow the definitions provided by Albuquerque et al. [2]: (i) *DSL Expressiveness*, the extent to which a DSL allows to directly represent the elements of a domain; and (ii) *DSL Conciseness*, the economy of terms without harming the artifact comprehension. We used Usa-DSL, a DSL usability taxonomy proposed by Poltonieri et al.[32–34] to define the usability evaluation metrics for our study. The evaluation metrics show a direct relationship to the questions posed on the questionnaire. In such a manner, we evaluate usage efficiency (Q1), expressiveness (Q2, Q5), flexibility (Q3), conciseness (Q4), satisfaction (Q6, Q12), ease of learning and use (Q7, Q8, Q10), and readability (Q9). We also included a question about the confidence of the participants in the solution implemented with DAMI-DSL (Q11). We also included one qualitative question (Q13), for descriptive answers, asking the participants to leave feedback regarding their experience with the DSL. These qualitative insights enriched our understanding of factors influencing the interactions of participants with DAMI-DSL.

The evaluation involved 21 participants selected through convenience sampling [4]. This group comprised eleven computer science students and ten developers from our research group's development team. To recruit participants, we invited students and developers from our work environment to voluntarily participate in the experiment. We specifically chose developers who had no prior knowledge of DAMI-DSL and were not related to this project in any way. In addition to the questionnaire handed out to the participants, we asked additional questions about their demographics and previous experience. The selected individuals, representing potential users of the DAMI-Framework were asked about their experience both in software development and data migration. Table 3 provides an overview of the participant's demographics. Participants were also inquired about their level of familiarity with scripting, DSLs, and SQL. Among the 21 participants, there where 15 males and five females. Experience in development ranges from less than one year to 3–5 years, with a majority (12 participants) having less than a year of experience. In terms of data migration experience, half of the participants (11) reported having no experience, while the others reported some experience. When it comes to scripting knowledge, participants exhibited a range of familiarity, from "Unfamiliar" to "Expert," with five reporting "Proficient" status. Regarding their exposure to DSLs, 16 participants indicated no prior experience, while five had some familiarity. Lastly, SQL proficiency varied widely, with seven participants identifying as "Proficient," 10 as "Competent," and four as "Novice." This distribution indicates a blend of backgrounds and expertise levels, contributing to a comprehensive evaluation of DAMI-DSL.

**Table 2: Questionnaire for the survey with developers**

| Goal | Evaluate the DAMI-DSL usability and limitations from the perspective of DSL properties | | |
|---|---|---|---|
| Q# | Metric | Question | Type |
| Q1 | Efficiency | How do you feel about the efficiency of the data migration tasks using the DSL? | 5-Point Scale |
| Q2 | Expressiveness | To what extent do you think the DSL allowed you to express your data migration logic clearly and effectively? | 5-Point Scale |
| Q3 | Flexibility | Were there limitations in expressing certain types of data migration tasks with the DSL? | 5-Point Scale |
| Q4 | Conciseness | How would you rate the conciseness of the DSL in comparison to GPLs you are familiar with, like SQL? | 5-Point Scale |
| Q5 | Expressiveness | How did you find the representation of data migration tasks by the DSL? | 5-Point Scale |
| Goal | Evaluate user satisfaction with DAMI-DSL | | |
| Q# | Metric | Question | Type |
| Q6 | Satisfaction | How satisfied are you with the DSL for data migration? | 5-Point Scale |
| Q7 | Ease of Use | How did you feel about learning and using the DSL for data migration tasks? | 5-Point Scale |
| Q8 | Ease of Use | How many difficulties or challenges have you encountered while using the DSL? | 5-Point Scale |
| Q9 | Readability | How clear and readable did you find the DSL code for the data migration tasks? | 5-Point Scale |
| Q10 | Ease of Use | How would you describe the learning curve of the DSL for someone with your level of experience in data migration? | 5-Point Scale |
| Q11 | Confidence | How confident are you with the solution you provided? | 5-Point Scale |
| Q12 | Satisfaction | Based on your experience, would you recommend the DSL for data migration tasks to your colleagues or peers? | 5-Point Scale |
| Goal | Evaluate the overall impressions of developers about DAMI-DSL | | |
| Q# | Metric | Question | Type |
| Q13 | Qualitative | Please provide any additional comments or feedback regarding your experience with the DSL for data migration. | Open-ended |

**Table 3: Participant Demographics and Experience**

| Occupation | | | | |
|---|---|---|---|---|
| **SW Developer** | | 10 | **CS Student** | 11 |
| **Gender** | | | | |
| **Male** | | 15 | **Female** | 5 |
| **Data Migration Experience** | | | | |
| **<1 yr.** | **1-2 yrs.** | | **3-5 yrs.** | **>5 yrs.** |
| 11 | 9 | | 1 | 0 |
| **Scripting Languages Knowledge** | | | | |
| **Unfamiliar** | **Novice** | **Competent** | **Proficient** | **Expert** |
| 0 | 1 | 7 | 8 | 5 |
| **SQL Expertise** | | | | |
| **Unfamiliar** | **Novice** | **Competent** | **Proficient** | **Expert** |
| 0 | 4 | 10 | 7 | 0 |
| **DSL Experience** | | | | |
| **Yes** | | 5 | **No** | 16 |

## 5.2 Study Results and Discussion

This section presents the results of the evaluation of DAMI-Framework. We begin by analyzing the responses to the 12 quantitative questions, each rated on a five-point Likert scale. Next, we examine insights from the qualitative question to gather general feedback on DAMI-DSL. The quantitative questions are categorized into two groups based on the specific aspects of DSL they measure. The first group addresses the properties of DAMI-DSL, while the second group focuses on user experience.

**DSL Properties:** The quantitative questions on DSL properties were designed to assess the efficiency, conciseness, expressiveness, and flexibility of DAMI-DSL. As shown in Figure 4(a), all 21 participants responded positively, with each participant selecting either "Agree" or "Strongly Agree" for all five questions related to DSL properties. This consistent approval resulted in a 99% positive response rate, demonstrating DAMI-DSL's effectiveness in achieving efficiency and conciseness. Responses to Q1 indicated that 100% of participants found DAMI-DSL to be "Efficient" or "Very Efficient". For Q2, all participants described it as "Expressive" or "Really Expressive". In Q3, 95% of responses indicated that the language had "No limitations", with only one participant remaining neutral. Finally, for Q4 and Q5, 100% of participants rated DAMI-DSL as "Much more concise than GPL" and "Very concise", respectively. The single neutral response regarding flexibility is understandable, given that DAMI-DSL, a domain-specific language, is inherently less flexible than a GPL and was explicitly designed for data migration tasks.

**User Experience:** To evaluate user experience, we assessed the ease of use, readability, and overall satisfaction with DAMI-DSL. As shown in Figure 4(b), participant feedback indicates that DAMI-DSL is generally perceived as accessible and straightforward. Based on responses from 21 participants, 91% reported either "Agree" or "Strongly Agree" regarding their overall satisfaction, while only 9% chose "Neutral", reflecting a largely positive impression. Qualitative feedback supports these findings; participants found the "*syntax of the DSL quite intuitive*" (P3, P16), attributing this to its resemblance to natural language, which makes it accessible for users with limited data migration experience. Other participants mentioned that

```
1  MAP publication, book TO paper (
2      pub_title TO title,
3      pub_id TO doi
4  ) WHERE (pub_id=book_id)
5  GET publication_code FROM publication.code WHEN
        id_publication=pub_id;
6  MAP publication, book_chapter TO paper (
7      pub_title TO title,
8      pub_id TO doi,
9      chapt_start_page TO startpage,
10     chapt_end_page TO endpage
11 ) WHERE (pub_id=chapt_id)
12 GET publication_code FROM publication.code WHEN
        id_publication=pub_id;
13 MAP publication, conference TO paper (
14     pub_title TO title,
15     pub_id TO doi,
16     conf_start_page TO startpage,
17     conf_end_page TO endpage
18 ) WHERE (pub_id=conf_id)
19 GET publication_code FROM publication.code WHEN
        id_publication=pub_id;
20 MAP publication, journal TO paper (
21     pub_title TO title,
22     pub_id TO doi,
23     jrnl_start_page TO startpage,
24     jrnl_end_page TO endpage
25 ) WHERE (pub_id=jrnl_id)
26 GET publication_code FROM publication.code WHEN
        id_publication=pub_id;
```

**Listing 5: Solution provided by participants**

DAMI-DSL code was "*much easier to understand and clear than a traditional GPL*" (P4, P10, P14, P18), as the code format facilitates a direct visual connection between database tables. Other participants appreciated the readability of the DSL, stating it "*supports a gentler learning curve, making it easier to introduce to individuals without prior experience in data migration*" (P17).

Participants praised DAMI-DSL for its efficiency and conciseness, noting its effectiveness in streamlining the data migration and simplifying tasks. Listing 5 shows an example of the code submitted by participants as the solution to the assigned task. All participants provided similar solutions for the DAMI-DSL task, with only minor variations. The most common errors observed were typing mistakes that occurred when copying and pasting code, often resulting in missed updates to attribute names. Only one participant made significant errors due to a misunderstanding of the task rather than issues related to the language.

Overall, responses with DAMI-DSL produced only eight errors across 21 participants, resulting in an error rate of approximately 0.38. For SQL, the error rate was about 1.57, with 33 errors in the same number of responses. This represents an error reduction of approximately 75.80%, demonstrating that DAMI-DSL significantly decreases errors compared to SQL in this experiment. These findings suggest that DAMI-DSL is robust for data migration tasks. Participants also appreciated how effectively DAMI-DSL conveyed migration logic, particularly in illustrating table connections. For instance, P2 and P4 stated that "*the DSL helps us visualize how the source and destination tables are directly connected in the code, making it easier to follow the migration process flow and understand what* is happening", and that "*with a more complex schema, the advantage of using this DSL would be much more noticeable*", respectively.

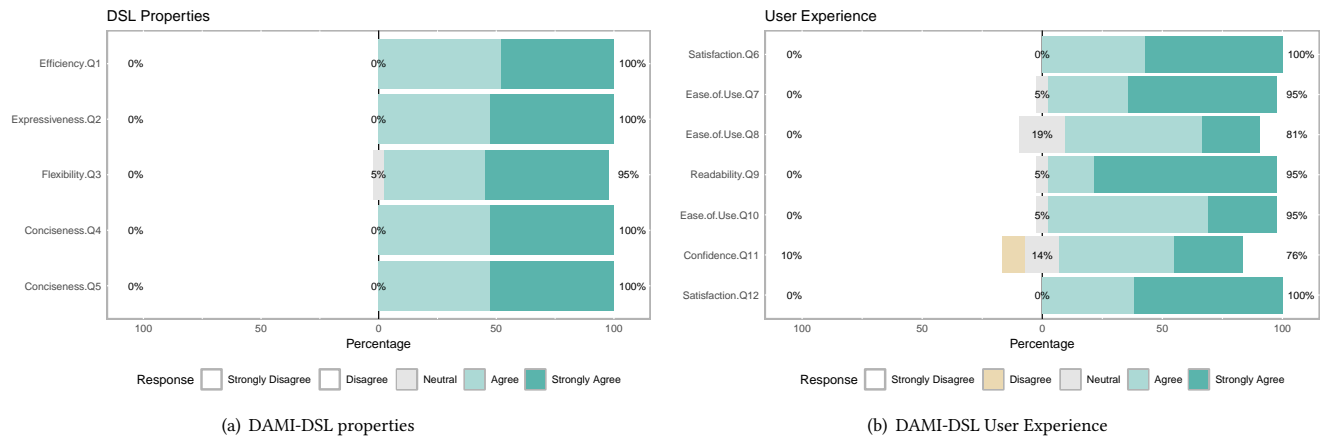

(a) DAMI-DSL properties

(b) DAMI-DSL User Experience

Figure 4: Overview of the software developers' perception when using DAMI-DSL.

We recorded each participant's time to complete the assigned tasks during the experiment. Participants were divided into two subgroups: one group completed the DSL tasks first, while the other started with SQL. Overall, the average time spent on the DSL tasks was similar to that of the SQL tasks. However, we noticed that participants took longer on their first task regardless of the programming language due to their initial exposure to the migration process, which required extra time to understand the domain. As participants P10 and P11 noted, "*the most challenging part of the experiment was understanding the domain rather than using the language for migration*". P11 elaborated further, saying, "*SQL seemed easier, but only because it was my second task. What really took me time was understanding the domain. Once I had done that and was rewriting it in the form, I realized that the DSL was much simpler and easier to understand*".

Additionally, participants with less experience in SQL and data migration showed a significant reduction in completion time when using DAMI-DSL for their second task. Specifically, students who began with DAMI-DSL took 12 minutes to finish the second task, while those who started with SQL completed the DSL task in just seven minutes. This indicates that DAMI-DSL may be faster for users with limited technical knowledge once they become familiar with the domain, highlighting the tool's potential to simplify migration tasks for non-technical users.

**Limitations and improvement opportunities:** Overall, feedback on DAMI-DSL was positive; however, some participants reported challenges when using DAMI-DSL. Although no participants provided negative responses, several noted difficulties at specific points in their experience. Notably, the question on confidence (Q11) received the lowest ratings, likely due to participants' uncertainty about ensuring their solution accuracy when working with an unfamiliar language. For example, P5, P9, and P18 expressed concerns about their confidence in the solutions they implemented with DAMI-DSL, suggesting a need for additional support or guidance when adopting the DSL. Potential improvements could include expanded documentation on DAMI-DSL syntax and the internal workings of the DSL Parser to facilitate a smoother adoption process.

Participants P1, P2, and P3 suggested that enhancing the readability of DAMI-DSL could involve simplifying how relationships between old and new primary keys are maintained. Analysis of qualitative data reveals that mapping legacy primary keys to newly generated ones is the most challenging aspect of the syntax. This operation is found on lines 5, 12, 19, and 26 of Listing 5. P1 described this operation as "*confusing*" and "*difficult to understand*", likely due to the high cognitive load required—often termed Hard Mental Operations—where developers must simultaneously navigate multiple notations and abstractions, a known usability challenge [15]. Additionally, P3 pointed out that integrating SQL functions into DAMI-DSL might seem counterintuitive for users who chose this DSL to distance themselves from other programming languages.

### 5.3 Threats to Validity

We discuss the threats to the validity of our study, and how we mitigate them, based on the taxonomy by Wohlin et al. [48].

**Construct Validity:** We employ a convenience sampling strategy [47], which may introduce a threat to population validity. To address this, we made sure that the person running the experiment was not a professor or superior to any participants, reducing the chance of a subject-expectancy effect [21]. We also gathered information about the participants' relevant experience and included this in our analysis. However, gathering this background data might have created a stereotype threat, making less experienced participants feel less confident when reminded of their inexperience. To mitigate this threat, we only collected demographic data after all tasks and surveys were completed. Lastly, the answers to our closed-ended questions could unintentionally bias the responses. To reduce this risk, we provided textual descriptions for each option on a Likert scale and included open-ended questions so participants could share their thoughts.

**Internal Validity:** A potential threat is that participants might not understand the instructions. To address this, the experimenter explained the tool and tasks, and provided an example, encouraging participants to ask questions if they were confused.

**Conclusion Validity:** The sequence of tasks could affect results because of the learning effect [39]. To reduce this threat, we organized the tasks using a counterbalancing approach [21]. Participants were divided into two groups that performed the tasks in different orders.

**External Validity:** Although our evaluation is based on a real-world case, it may not capture the complexities of actual data migration fully. Participants' experiences and perceptions could differ significantly when handling more intricate and diverse datasets or migration requirements. Our evaluation focuses on a single practical scenario, which offers a detailed view of our current project but limits the generalizability of our findings. To address this limitation, we plan to apply our framework to additional use cases, including scenarios in digital libraries and geographic information systems, to further test and validate our solution.

## 6 Related work

Data migration has long been a popular study topic, with various methodologies being investigated [11]. Next, we discuss related work targeting data migration in the context of legacy systems' modernization, migration, re-engineering, or reverse engineering.

Hudicka [18] proposed an approach in which the database migration is split into seven stages. However, this author does not present a detailed solution to face practical challenges, such as the ones presented in Section 2. Elamparithi et al. [12] review existing migration strategies in Relational Database Migration (RDBM) and conclude that existing techniques do not provide a solution suitable for more than one target database or are unable to manage both schema and data conversion. Maatuk et al. [22, 23] proposed a solution for RDBM that takes an existing relational database as input, enriches its metadata representation with as much semantics as possible, and constructs an enhanced Relational Schema Representation (RSR). Based on the RSR, a canonical data model is generated, which captures essential characteristics of the target data models suitable for migration. Similarly, Wider [46] presents a generated domain-specific workbench for the nanophysics systems and a taxonomy of synchronization types that allows defining which model transformations are required for view synchronization in that workbench.

Other studies have employed Model-Driven Engineering (MDE) techniques to address data migration issues. Bermudez et al. [40] employed MDE techniques to enhance the quality of a relational database automatically. Their work automates the discovery of foreign keys and functional dependencies in existing relational databases, which is then used to normalize the database. Terzic et al. [44] presented a DSL for designing NoSQL validation schemas, which was used as part of a data migration tool for migrating relational databases to document-oriented databases. Oubelli et al. [27] addressed the management of multiple versions of data models for space missions using MDE techniques. Their work consists of a set of operators to encapsulate the intricacies of schema and data migration, guaranteeing that data is preserved correctly across multiple versions of the data models.

Overeem et al. [28] considered migration triggers, microservices architecture, and event sources as technological solutions for automated data migration in MDE-developed systems. Aboulsamh et al. [1] also address software evolution and data migration from an MDE perspective. They extended MDE approaches to support data migration. For instance, UML represents changes as a model, which is then applied as the basis for automatically generating corresponding data migration scripts in SQL.

There are plenty of solutions for schema evolution and tools that provide support for the ETL processes, but they do not directly solve the problem we investigate in our work. Our data migration should enable the schema migration so that the data can fit into the new schema. However, most of the solutions studied in the literature review only support data cleaning and transformation within the same schema or, given the possibility to change the schema, the operations provided are too limited for our use cases [9].

Although the work presented above addresses specific challenges in data migration of generic software systems, with solutions produced using RDBM or MDE techniques, they do not provide a general solution for specifying the many actions that must be performed during a data migration process. DAMI-Framework fills this gap with a flexible solution to support seamless data migration between database schemas during modernization.

## 7 Concluding Remarks

This paper describes the challenges faced in practice to migrate data as part of software modernization, which are mainly about *what* to migrate and *how* to retrieve and transform data. To tackle these challenges, we developed the DAMI-Framework, featuring a DSL to simplify migration script creation and a parser for automatic SQL script generation. Our evaluation shows that the DAMI-Framework effectively reduces migration efforts, provides a strong language for schema mappings, and delivers a positive user experience.

The outcomes of our empirical assessment shed important light on how well DAMI-DSL works and how user-friendly it is for data migration activities. Although the participants highly praised the DSL for its readability, efficiency, and ease of use, they also suggested areas that needed work, such as addressing difficulties encountered and improving readability. DAMI-DSL will be improved and refined in response to user feedback. After these improvements, future work will include making the DAMI-Framework available as an open-source solution.

## 8 Data Availability

Artifacts of the evaluation are available online [38].

## Acknowledgments

CITIC, as a center accredited for excellence within the Galician University System and a member of the CIGUS Network, receives subsidies from the Department of Education, Science, Universities, and Vocational Training of the Xunta de Galicia. Additionally, it is co-financed by the EU through the FEDER Galicia 2021-27 operational program (Ref. ED431G 2023/01); GRC: ED431C 2021/53, GAIN/Xunta de Galicia; TED2021-129245B-C21: MCIN/AEI/10.13039/501100011033 and "NextGenerationEU"/PRTR; PID2021-122554OB-C33: MCIN/AEI/10.13039/501100011033 and EU/ERDF A way of making Europe; PID2022-141027NB-C21: MCIN/AEI/10.13039/501100011033 and EU/ERDF A way of making Europe; PRE2021-099351: MCIN/AEI/10.13039/501100011033 and "FSE+Fondo Social Europeo Plus".